\documentclass[a4paper,11pt]{article}
\usepackage[left=3cm,right=3cm,
top=3cm,bottom=3cm]{geometry}
\usepackage[T2A]{fontenc}
\usepackage[utf8]{inputenc}
\usepackage{authblk}
\usepackage[english]{babel}
\usepackage{amsmath,amssymb,amsthm,mathtools}
\usepackage{euscript,mathrsfs}
\usepackage{icomma}
\usepackage{graphicx}
\usepackage{wrapfig}
\usepackage[dvipsnames]{xcolor}
\usepackage{indentfirst}
\usepackage{amsbsy}
\usepackage{wasysym}
\usepackage{cancel}
\usepackage{verbatim}
\usepackage{empheq}
\usepackage{adjustbox}
\usepackage{cite}
\usepackage{physics}
\usepackage[pdfencoding=auto]{hyperref}
\definecolor{urlcolor}{HTML}{120099}
\definecolor{linkcolor}{HTML}{005F5F}
\hypersetup{pdfstartview=FitH,  linkcolor=linkcolor,urlcolor=urlcolor, colorlinks=true,citecolor=blue}

\usepackage[framemethod=tikz]{mdframed}
\newmdenv[innerlinewidth=0.5pt, roundcorner=4pt,linecolor=blue,innerleftmargin=6pt,
innerrightmargin=6pt,innertopmargin=6pt,innerbottommargin=6pt]{mybox}

\usepackage{tikz} 
\usetikzlibrary{calc,through,backgrounds,patterns,decorations.pathmorphing,decorations.markings, trees,positioning,arrows,arrows.meta, chains,shapes.geometric,%
    decorations.pathreplacing,shapes,arrows}  
\tikzset{->-/.style={decoration={markings,mark=at position #1 with {\arrow{>}}},postaction={decorate}}}

\renewcommand{\phi}{\varphi}

\title{On Schwinger-like pair production of baryons and new non-perturbative processes in electric field}
\author[b]{Alexander Gorsky }
\author[a,b]{Arseniy Pikalov}

\affil[a]{Moscow Institute of Physics and Technology, Dolgoprudny 141700, Russia}
\affil[b]{Institute for Information Transmission Problems, Moscow 127994, Russia}

\begin{document}

\maketitle
\begin{abstract}
    We consider the  Schwinger production of baryons in an external
    electric field in the worldline instanton approach. The process occurs in the confinement regime 
    hence the holographic QCD and the Chiral Lagrangian are used as the tools. The new exponentially 
    suppressed processes in a constant electric field involving the composite worldline 
    instantons are suggested. These
    include the non-perturbative decay of a neutron into a proton and charged meson and the 
    spontaneous production
    of $p\bar{n}\pi^{-}$ and $n\bar{p}\pi^{+}$ states.
\end{abstract}

\section{Introduction}
The instanton solutions with finite action  describe 
one or another tunneling phenomena and are 
well localized in the Euclidean space-time. Another example of 
tunneling phenomena corresponds 
to the  bounces which  can be thought of approximately as 
the instanton-antiinstanton pair. The familiar examples 
are the classical Euclidean solutions corresponding to the Schwinger charged pair 
production in the electric field \cite{affleck1982pair} or spherically 
symmetric bounce for a false vacuum decay. The solutions have finite action
and negative mode in the spectrum of fluctuations which indicates the instability of the ground state
in the Minkowski space-time. For the Schwinger pair production
the bounce can be interpreted as the worldline trajectory of a point-like particle \cite{affleck1982pair}
and the standard perturbation theory can be applied to justify
the dominance of the leading order in the perturbation theory.

A more delicate situation occurs when we consider the monopole pair production
in the external magnetic field \cite{affleck1982monopole} since a monopole is not 
a fundamental particle. However,
the worldline instanton approach works reasonably well and the probability
of the monopole pair production can be derived in a weak magnetic 
field \cite{affleck1982monopole}.  Other interesting processes that can be 
discussed in the worldline instanton framework are the  
monopole decay in the external electric field and decay of the electrically
charged particle in the magnetic field \cite{gorsky2002schwinger,monin2005monopole}. In these
examples the worldline instantons are composite. The approach works perfectly
for the inhomogeneous external fields as well \cite{dunne2005worldline}.

The worldline instanton approach is suitable for discussion of the Schwinger
process in the strong coupling regime. The evaluation of the Wilson
or t'Hooft loops in Euclidean space-time representing worldline instantons at strong coupling 
can be performed by holographic means and gets reduced to the calculation
of the area of a minimal surface in some gravitational background. The holographic Schwinger effect has been discussed
in the $N=4$ SYM in \cite{gorsky2002schwinger,semenoff2011holographic,ambjorn2012remarks,sonner2013holographic}
and has been extended for non-conformal backgrounds in \cite{sato2013holographic,kawai2015holographic, ghodrati2015schwinger,grieninger2023entanglement2,grieninger2023entanglement} for the
creation of the massive quark-antiquark pair. The evaluation 
of the Wilson loop corresponds to the Schwinger process for the 
electric particles while the t'Hooft loop to the
creation of the magnetically charged pair. The minimal surface mimics the  effects of  all-order
perturbation theory.

In the confinement regime, 
the most suitable approaches for descriptions of baryons are holographic QCD and Chiral Perturbation Theory (ChPT).
In the 
ChPT the baryon is identified as the topological classical solution to the equations of motion
stabilized by the Skyrme term \cite{witten1979baryons}. More
recently the pancake configuration for $N_f=1$ baryons has been
found \cite{komargodski2018baryons,karasik2020skyrmions}.
Holographically baryon is 
represented by the baryonic vertex -- D4 brane wrapped around $S^4$ with $N_c$ attached
strings \cite{witten1998baryons}. Some other aspects of the supergravity
aspects of the baryon vertex were discussed in \cite{gross1998aspects,callan1999baryons,brandhuber1998baryons}.
In the simplified hard-wall holographic QCD
model baryon can be identified with an instanton solution in the 
5d gauge theory on the flavor branes \cite{son2004qcd}.
In more realistic Witten-Sakai-Sugimoto (WSS) model \cite{witten1998anti,sakai2005low} or in some other
versions of holographic QCD baryons have been 
considered in \cite{hata2007baryons,hashimoto2008holographic,seki2009comments,dymarsky2011attractive,hong2007dynamics,sonnenschein2016,kim2008electromagnetic,sutcliffe2010skyrmions}(see \cite{jarvinen2022holographic} for the
recent review). If the chiral condensate is taken into account for the holographic baryon 
the dyonic instanton solution in 5d gauge theory becomes relevant \cite{gorsky2012baryon,gorsky2015baryon}.

In this study, we consider the Schwinger production of the baryon pair 
in the worldline instanton approach. Since baryon is composite the situation
is somewhat similar to the monopole pair production.
In the ChPT we shall find the worldline instanton for a Skyrmion anti-Skyrmion pair  and
assume that the radius of the Skyrmion Euclidean worldline circle 
is large enough to neglect the loop corrections via meson exchanges.
In the holographic WSS model we have two contributions to the effective action, one comes from the 
worldline circle of instanton-antiinstanton pair  in 5d gauge theory  
or the baryonic vertex-antivertex pair in the stringy terms.
The second contribution comes from the worldsheet action of $N_c$
strings attached to the baryonic vertex which end at the flavor D8 branes. The contribution of the baryon vertex
in a weak electric field comes from the action evaluated at the circular classical trajectory and
is more or less parallel to the ChPT calculation.
However, the effect from the $N_c$-strings is more tricky since we have 
a vertex-anti-vertex pair and the strings attached 
to the vertex can end either at the flavor branes or at anti-vertex. We shall comment
on both possibilities.

We shall also suggest new non-perturbative processes for baryons in the 
external electric field which are suppressed exponentially. All processes will
be described by  the composite worldline instantons in Euclidean space-time 
which involve several segments 
of worldlines of the particles in the external field.
One type of such composite worldline instantons yields the
channel of the neutron decay into the proton and $\pi^{-}$ meson  in the constant electric 
field somewhat analogously to the
decay of the monopole in the external electric field considered in 
\cite{gorsky2002schwinger,monin2005monopole}. Another type of the composite 
worldline instantons will provide the process of spontaneous creation of
$p\bar{n}\pi^{-}$ and $n\bar{p}\pi^{+}$ final states. Such type of 
composite worldline instantons have been considered in the cosmological 
context in \cite{gorsky2000tunnelling,gorsky2007spontaneous}.

The paper is organized as follows. In Section 2 we remind the worldline instanton method 
and evaluate the probability of the proton-antiproton pair production in the homogeneous external electric field in ChPT. 
In Section 3 we discuss the Schwinger process in the holographic
QCD and focus on the differences with the ChPT calculation due
to the strings attached to the baryonic vertex. In Section 4
we suggest the new non-perturbative processes involving the composite worldline
instantons and evaluate their rates in the leading approximation. We discuss the key findings and open
questions in the Conclusion. In Appendix A we present some formulas concerning the dyonic
instanton solution for baryon which takes into account the chiral condensate.
In Appendix B we consider the example of holographic evaluation of the Schwinger process rate when
the baryon involves a heavy quark.

\section{Worldline instantons and baryon pair creation}

\subsection{Worldline instantons for Schwinger processes}

Let us recall the worldline approach for the Schwinger pair production 
of massive particles in the 
constant external electric field. 
We perform the Wick rotation to the Euclidean space and consider the classical Euclidean  trajectory minimizing the action
which involves the contribution proportional to the length of the space-time trajectory and 
the term from the interaction with the external electric field
\begin{equation}
    S= m\int d\tau \sqrt{\dot{x}^2} +i\oint A .
\end{equation}
In Euclidean space-time 
the electric field turns into the magnetic field hence from the Euclidean viewpoint the 
trajectory of the massive particle  is the solution to the
equation of motion 
\begin{equation}
    m\frac{d}{d\tau}(\frac{\dot{x}_{\mu}}{\sqrt{\dot{x}^2}})=iF_{\mu\nu}\dot{x}_{\nu},
\end{equation}
that is the Larmour circular trajectory.
The corresponding effective action for the radius of trajectory reads as
\begin{equation}
S = 2 \pi r_0 m -  q\pi r_0^2 E, 
\end{equation}
where $m$ is the mass of the particle. The second term is proportional to the flux through the Euclidean trajectory.
The action extremization yields 
\begin{equation}
r_0= \frac{m}{qE}, \quad S = \frac{\pi m^2}{q E}.
\end{equation}
and the leading exponential contribution to particle production rate per 
unit volume is proportional to $\Gamma \sim e^{-S}$.

The full result which is derived via the evaluation of the imaginary part of the effective action 
in the external field can
be derived in the worldline approach as well. To this aim, one has to take into account the multiple 
winding trajectories which yield $e^{-nS}$ contributions, evaluate the determinant of the quadratic
fluctuations at the top on the classical trajectories and sum up over all $n$. This procedure exactly
reproduces the weak coupling result for the fermion pair production \cite{affleck1982pair}
\begin{equation}
    \omega=\frac{(eE)^2}{4\pi^2} \sum_{n=1}^{\infty}\frac{1}{n^2} \exp(-\frac{n\pi m^2}{eE})
\end{equation}
The first perturbative
correction to the action at the classical trajectory due to the photon Coulomb exchange can be easily
evaluated as well \cite{affleck1982pair}. The standard tools are not effective for the evaluation of the higher-order 
corrections to the classical action that is why the holographic approach turned out to be useful 
\cite{gorsky2002schwinger,semenoff2011holographic,ambjorn2012remarks}.

In this study, we shall apply the worldline approach for the evaluation of the baryon pair production 
in the external electric field.  We would like to consider the approximation when a baryon is 
a point-like solitonic object whose size and mass are under control.  
In the leading approximation, we evaluate the action on the
circular worldline instanton trajectory. The leading result is subject to corrections
of the different types. We assume that the radius of the instanton is much larger than
 $M_p^{-1}$, which allows us to disregard the polarizability of the nucleon in the external electric field.
At large radius, we can neglect the nucleon-antinucleon interactions due to massive meson exchanges.
The electromagnetic   Coulomb interaction between the nucleon-antinucleon pair is small.

\subsection{Baryon pair production  in the electric field}

Consider the Chiral Lagrangian which without external fields in Euclidean time $\tau$ looks as follows
\begin{equation}
  L = \frac{f_\pi^2}{4} Tr\left(\partial_\mu U \partial_\mu U^\dagger \right) + \frac{1}{32 e^2} Tr\left( \left[U^\dagger\partial_\mu U ,U^\dagger \partial_\nu U \right]^2\right) .
\end{equation} \label{skyrme}
When coupled  to the external $U(1)$ gauge field
the baryonic current reads as \cite{witten1979baryons,callan1984monopole}
\begin{equation}
\label{baryonic}
 B^\mu = \frac{i}{24\pi^2} \epsilon^{\mu \nu \sigma \rho}Tr(L_\nu L_\sigma L_\rho) + 
 \frac{\epsilon_{\mu\nu\alpha\beta}}{24 \pi^2}\partial_{\nu}[3ieA_{\alpha}Tr Q(U^{-1}\partial_{\beta} U
 +\partial_{\beta}U U^{-1}],
   \end{equation}
  where  $L_\mu = U \partial_\mu U^\dagger$.
The electromagnetic current has the following form
\begin{equation}
    J^Q_{\mu}= J^3_{\mu} + \frac{1}{16\pi^2} \epsilon_{\mu\nu\alpha\beta}  Tr[ Q\partial_{\nu} UU^{-1}\partial_{\alpha} U U^{-1}\partial _{\beta} 
    UU^{-1} + U^{-1}\partial_{\nu} U U^{-1}\partial_{\alpha} UU^{-1}\partial_{\beta} U ]
    \end{equation}
    $$
    + \frac{ie}{4\pi^2}\epsilon_{\mu\nu\alpha\beta}\partial_{\nu} A^{\alpha} Tr[Q^2 \partial_{\beta} U U^{-1}
    U^{-1}\partial_{\beta}U + Q\partial_{\beta}UQU^{-1} -\frac{1}{2} QUQ\partial_{\beta}U^{-1}].  
$$
The anomalous term in the baryonic current is responsible for the Callan-Rubakov effect of 
monopole catalysis of the proton decay in  the Skyrme model \cite{callan1984monopole} while
anomalous term in the electromagnetic current is responsible for the $\pi^0\rightarrow 2\gamma$ decay
process.

The Skyrmion solution to the equations of motion can be written in the following ansatz
\begin{equation}
   U_0 = e^{i f(r) \hat{x} \vb*{\sigma}},\; f(0) = \pi, \; f(\infty) = 0. 
\end{equation}   
where the exact form of $f(r)$ can be found by numerical minimization of the action  \eqref{skyrme}. We are interested in the low-energy dynamics of the baryon in the presence of external electromagnetic field. To obtain the corresponding effective action we introduce the collective coordinates $C(t)$  and $x_c(t)$ :
\begin{equation}
    U(x, t) = C(t) U_0(x - x_c(t)) C(t)^{-1}.
\end{equation}
and substitute this ansatz into the Chiral Lagrangian.

Let us consider the  electric field in $z$ axis direction and
\begin{equation}
    A_t = - E z, \quad A_i = 0,
\end{equation}
hence the only non-vanishing  strength component is
\begin{equation}
    F_{tz} = -F_{zt} = E.
\end{equation}
In this case, the anomalous electromagnetic current does not contribute to the action while the first term yields
\begin{equation}
    L_a = \frac{e}{16 \pi^2} A_0  \varepsilon^{0 \nu \alpha \beta} \Tr Q 
     \left( \partial_{\nu} U U^{-1} \partial_{\alpha} U U^{-1} \partial_{\beta} U U^{-1} + U^{-1} \partial_{\nu} U U^{-1} \partial_{\alpha} U U^{-1} \partial_{\beta} U \right),
\end{equation}
The trace is proportional to the Skyrmion topological charge density and after integration over spatial coordinates we have
\begin{equation}
    \int d^4 x L_a = \int dt \int 4 \pi r^2 dr \frac{e}{16 \pi^2} \frac{4}{r^2} (\sin^2 f) f' A_0 = \int dt \frac{e n}{2} A_0. 
\end{equation}
This is a standard expression for the action of a charged particle in electric field where $n$ is Skyrmion topological charge.

The matrix $C(t)$ parametrizing the internal rotation can be chosen in the following form  $C = a_0 + i \vb*{a} \vb*{\sigma}$ ($\vb*{\sigma}$ -- vector of Pauli matrices).
If $C(z)$ changes smoothly enough, the moduli space action is
$$
S = M L + \Lambda\int dz (a_0')^2 +(\vb*{a}')^2, \; a_0^2 + \vb*{a}^2 = 1,
$$
therefore the low-energy dynamics is that of a particle on $S^3$.
The constant $\Lambda$ can be expressed as integral of the profile function and 
the wave functions with definite angular momentum are homogeneous polynomials in $a$ variables.
We are interested in the wave functions of proton and neutron found in \cite{witten1979baryons,witten1983global}:
$$
\ket{p \uparrow} = \frac{1}{\pi}(a_1 + i a_2), \ket{p \downarrow}  = -\frac{i}{\pi}(a_0 - i a_3);
$$
$$
\ket{n \uparrow} = \frac{i}{\pi}(a_0 + i a_3), \ket{n \downarrow}  = -\frac{i}{\pi}(a_1 - i a_2).
$$

Since in the leading approximation the effective action reduces to the action of the massive charged particle the worldline instanton solution looks
as before. Hence
the leading exponential factor for the proton-antiproton pair production 
in electric field evaluated at the Skyrmion worldline instanton reads as 
\begin{equation}
    w\propto \exp(-\frac{\pi M_p^2}{eE})
\end{equation}
This leading result is subject to the several types of corrections. The corrections due to the
higher winding numbers yield the expression $ \omega = \sum_{n} b(n)\exp(-\frac{n \pi M_p^2}{eE}) $
where $ b(n)$ takes into account the quadratic fluctuations at the top of the solution.
The corrections from the higher windings are small at weak electric field. The second type
of corrections comes from the meson exchange between the proton and antiproton. They 
involve the typical factor $\exp(-m_{mes}r)$ where $m_{mes}$ is a mass of the lightest 
meson  \cite{kang2014antinucleon}. We shall assume the condition $m_{mes}R_0>>1$, where $R_0$ is the radius of worldline instanton,
which suppresses the corrections
due to the meson exchange.

There is another possible source of correction to the probability rate.
It was noted in \cite{gorsky2020flavored} that there are also the S-Skyrmions localized 
in the Euclidean time which are analog of the S-brane in the string theory \cite{gutperle2002spacelike}.
Such an instanton solution is extended in some spatial direction, for example along the $z$ axis. In this case the profile function depends on  $r = \sqrt{x^2 + y^2 + \tau^2}$, $\hat{x} = (x, y, \tau)/r$.
The full baryonic current in the $z$ direction localized at Euclidean time reads as 
$$
J^B_\mu = \int dx dy d\tau B^\mu \propto \delta _{\mu z}.
$$
The scale of localization in the Euclidean time is fixed by the radius of the instanton.
These S-Skyrmions can be attached to the baryonic loop providing a more general
bounce configuration. In our study, we shall neglect these corrections.

\section{Baryon pair creation in holographic QCD}
In this Section, we are going to calculate the rate of baryon pair production
in WSS model of holographic QCD \cite{witten1998anti,sakai2005low}.
In the WSS model \cite{witten1998anti,sakai2005low} model at $T=0$ the holographic background 
looks as the cigar-like geometry involving coordinates $(r,\phi)$ supplemented with sphere $S^4$ and
four-dimensional Minkowski space-time. 
The flavor degrees of freedom
are introduced by adding $N_f$ $ D8-\bar{D8}$ branes extended along
all coordinates but $\phi$. The theory on the flavor D8 branes 
upon the dimensional reduction on $S^4$ yields the 5-dimensional Yang-Mills
theory with $SU(N_f)_R\times SU(N_f)_L$ gauge group supplemented with
the Chern-Simons term. The action reads as 
\begin{equation}
S= \sigma\int d^4x dz ( h(z) \Tr F_{\mu\nu}^2 +g(z) \Tr F_{\mu z}^2) + S_{CS}
\end{equation}
where $\mu,\nu = 1,2,3,4$ the metric factors are
\begin{equation}
h(z)=(1+z^2)^{1/3} \qquad g(z)=(1+z^2)
\end{equation}
and $\sigma$ is expressed through the t'Hooft coupling $\lambda$
as $\sigma= \frac{\lambda N_c}{216\pi^3} = a \lambda N_c$.
It yields the ChPT in the 
conventional low-energy QCD and  reasonable values of the
low-energy parameters. 

The baryon in the WSS model is identified as the D4 brane wrapped 
around $S^4$ and extended in the time direction. In terms of the 
5d YM theory with the flavor gauge group the baryon is the instanton solution localized 
in $(z,x_1,x_2,x_3)$
coordinates. Consider for example $N_f=2$ case and separate the $U(2)$ flavor gauge
field on D8 branes into the $SU(2)$ field $A(x,z)$ and U(1) field $B(x,z)$.
The solution for the instanton sitting around $(x=0,z=0)$ reads as 
\begin{equation}
A_{\mu}= -if(\eta)g_{inst}(x,z)\partial_{\mu}g_{inst}, \qquad A_0(x,z)=0, \qquad f(\eta)= \frac{\eta^2}{\eta^2 +\rho^2},
\end{equation}
where 
$$g_{inst}=  \frac{(z-z_0) -    i(\vec{x} - \vec{x_0})\vec{\tau}}{\sqrt{(z-z_0)^2 + |\vec{x} - \vec{x_0}|^2}}, $$
\begin{equation}
B_i(x,z)=0, \qquad B_0(x,z)= -\frac{1}{8\pi^2 \lambda \eta^2} [ 1 - \frac{\rho^4}{(\eta^2 +\rho^2)^2}],
\end{equation}
and we have introduced the notation:
\begin{equation}
\eta = \sqrt{(z-z_0)^2 + |\vec{x} - \vec{x_0}|^2},
\end{equation}
$\rho$  is a parameter, playing the role of the instanton's size. 
This solution is nothing but the Skyrmion solution 
and it realizes the
old Atiyah-Manton interpretation  \cite{atiyah1989skyrmions,eto2005skyrmions}.
The BPST instanton  can be used  as a good approximation since the solution 
is mainly localized around $z=0$
where the wrap factor can be neglected. 
The radius of the instanton solution in $(x,z)$ space is fixed at the extremum
of the corresponding potential
\begin{equation} \label{potential}
V(\rho)\propto \frac{\rho^2}{6} + \frac{1}{320\pi^2 a^2\rho^2}   \qquad \rho_{inst} = \frac{1}{8\pi^2 a}\sqrt{6/5}
\end{equation}
The second term in the potential comes from the Coulomb interaction due to CS term.
The baryonic charge $B$ gets identified as 
\begin{equation}
B=\int d^3 xdr( \Tr F_{L}\tilde{F}_{L} - \Tr F_{R}\tilde{F}_{ R}) 
\end{equation}
where $r$ is proportional to $z$. The position of the instanton is determined dynamically and it turns out that
it sits at the tip of the cigar.

Equivalently holographic baryon can be considered as the baryonic 
vertex with the $N_c$ string attached \cite{witten1998baryons}.
The baryonic vertex is the D4 brane wrapped around the internal $S_4$ part of the geometry. 
The mass of the baryon involves the contribution from the vertex as well as from the attached
strings. In the conventional picture, the baryonic vertex is placed at the tip of the cigar 
and the short strings attached to the vertex end at the flavor branes yielding $N_c$ quarks
in the antisymmetric representation.  Since the strings for massless quarks are short they do not contribute
to the mass of the baryon however if the baryon involves the heavy quark there is an additional contribution
from the string extended from the vertex to the corresponding flavor D8  brane.

Since we consider the baryon-antibaryon pair we encounter the new 
situation when some part of $N_c$
strings can connect instanton and antiinstanton. We shall consider three different situations
when all strings from vertex are short when all strings connect the instanton and antiinstanton
and the case when only part of the strings are stretched between the instanton-antiinstanton pair.

\begin{itemize}
    \item 

{\bf All strings are short}

In this case, the situation is similar to the Skyrmion bounce solution in ChPT considered above. 
The full bounce solution action is the sum of the instanton action in the electric field and the Nambu-Goto (NG) action for the strings.
\begin{equation}
    S = 2 \pi r M - q E \pi r^2 + S_{NG},
\end{equation}
    \begin{equation}
S_{NG} = T_{str} \int d\sigma \sqrt{det(g)},
\end{equation}
where $T_{str}$ is string tenstion,
 $M$ is the vertex mass, and $r \gg 1/M$ is the radius of the instanton loop which is localized at the tip of the cigar in the radial coordinate. 
The vertex-antivertex pair is 
 created and extended along the circular trajectory and we neglect the Nambu-Goto contributions 
 from short strings in this case, see Fig.~\ref{fig:stringstoD8}. If the circle is large we neglect the meson exchanges between emerging baryons. In the holographic setting the nucleon-antinucleon interaction via 
meson exchange has been discussed in \cite{kaplunovsky2011searching,kim2009nucleon}.

\begin{figure}[h]
\centering
\begin{tikzpicture}
\draw (-3, -2) -- (3, -2) -- (4, 2) -- (-2, 2) -- cycle;
\draw[dashed] (0.5,0) ellipse (2cm and 1cm);
\draw[fill=black] (-1.5, 0) circle (2pt);
\draw[fill=black] (2.5, 0) circle (2pt);
\draw[semithick] (-1.5, 0).. controls (-1.9, 0.3) ..(-2.3, 0);
\draw[semithick] (-1.5, 0).. controls (-1.8, 0.0) ..(-2.1, -0.3);
\draw[semithick] (-1.5, 0).. controls (-1.3, 0.0) ..(-1.0, -0.3);
\draw[semithick] (-1.5, 0).. controls (-1.7, 0.6) ..(-2.1, 0.4);
\draw[semithick] (-1.5, 0).. controls (-1.7, -0.3) ..(-1.5, -0.7);
\draw[semithick] (-1.5, 0).. controls (-1.3, +0.3) ..(-1.0, 0.4);
\draw[semithick] (-1.5, 0).. controls (-1.0, +0.3) ..(-0.8, -0.1);
\draw[semithick] (2.5, 0).. controls (2.9, 0.3) ..(3.3, 0);
\draw[semithick] (2.5, 0).. controls (2.8, 0.0) ..(3.1, -0.3);
\draw[semithick] (2.5, 0).. controls (2.3, 0.0) ..(2.0, -0.3);
\draw[semithick] (2.5, 0).. controls (2.7, 0.6) ..(3.1, 0.4);
\draw[semithick] (2.5, 0).. controls (2.7, -0.3) ..(2.5, -0.7);
\draw[semithick] (2.5, 0).. controls (2.3, +0.3) ..(2.0, 0.4);
\draw[semithick] (2.5, 0).. controls (2.0, +0.3) ..(1.8, -0.1);
\node at (3, 1.7){$D8$};
\node at (-1.5, 0.8) {$B$};
\node at (2.5, 0.8) {$\overline{B}$};
\end{tikzpicture}
 \caption{All strings are connected to D8 branes}\label{fig:stringstoD8}
\end{figure}
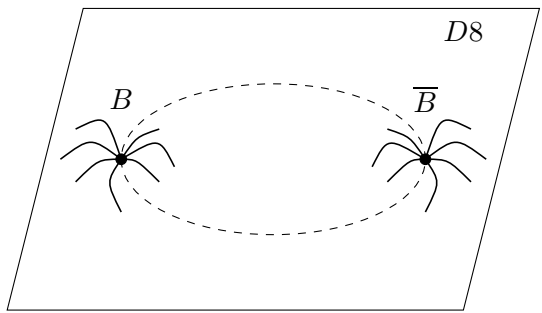

\item

{\bf Part of the strings are short}

The next option is as follows - some of the strings  attached to the vertex that end at anti-vertex are long
while the rest of the strings are short and  end at the flavor branes, see Fig.~\ref{fig:stringstoD8aninst}. 
In this case, we have to take into account the Nambu-Goto action for the long strings.
The first question concerns the geometry of the stringy worldsheet. The simple analysis
shows that the saddle point worldsheet configuration is the disk hence the action is 
\begin{equation}
    S = 2 \pi r M - q E_{eff} \pi r^2,
\end{equation}
where 
\begin{equation}
    E_{eff}=E - kT_{srt},\qquad T_{str}\propto \Lambda^2,
\end{equation}
if $k$ strings are long. The production rate in the leading approximation reads as
\begin{equation}
    \omega\propto \exp(- \frac{\pi M^2}{eE_{eff}}).
\end{equation}
Note that in the standard holographic framework, we assume the large $N_c$ limit
hence the probability is strongly suppressed as $ \exp(-N_c^2)$ 
since the baryon mass is proportional to $N_c$.
The process is possible only when $E_{eff}>0$ and is suppressed
with respect to the main channel of baryon pair production.

The key question concerns
the formation of the color singlet in the final state case. In the case of only one long string, the color singlet is possible and the emerging state upon the breaking of the string 
looks like the creation of a pair of baryon-antibaryon when a baryon is
built from the quark and the group of  $(N-1)$ quarks. The similar holographic configuration 
of baryon build from quark and diquark 
for $N_c=3$ has been discussed in \cite{sonnenschein2016}. Similarly when a singlet state is possible  
for $k>1$ more complicated baryon-like state built from two groups of quarks can be created.

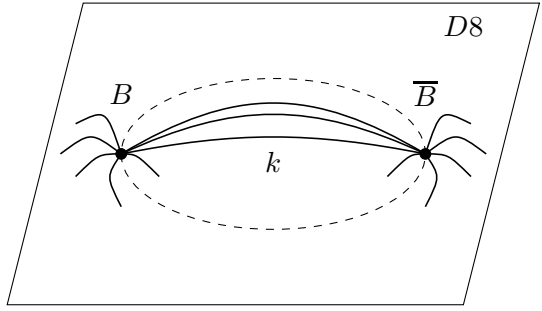
\begin{figure}
\centering
\begin{tikzpicture}
\draw (-3, -2) -- (3, -2) -- (4, 2) -- (-2, 2) -- cycle;
\draw[dashed] (0.5,0) ellipse (2cm and 1cm);
\draw[fill=black] (-1.5, 0) circle (2pt);
\draw[fill=black] (2.5, 0) circle (2pt);
%\draw[semithick] (-1.5, 0).. controls (0, 0.5) and (1, 0.5) ..  (2.5, 0);
\draw[semithick] (-1.5, 0).. controls (0, 0.3) and (1, 0.3) ..  (2.5, 0);
\draw[semithick] (-1.5, 0).. controls (0, 0.7) and (1, 0.7) ..  (2.5, 0);
\draw[semithick] (-1.5, 0).. controls (0, 0.9) and (1, 0.9) ..  (2.5, 0);
\draw[semithick] (-1.5, 0).. controls (-1.9, 0.3) ..(-2.3, 0);
\draw[semithick] (-1.5, 0).. controls (-1.8, 0.0) ..(-2.1, -0.3);
\draw[semithick] (-1.5, 0).. controls (-1.3, 0.0) ..(-1.0, -0.3);
\draw[semithick] (-1.5, 0).. controls (-1.7, 0.6) ..(-2.1, 0.4);
\draw[semithick] (-1.5, 0).. controls (-1.7, -0.3) ..(-1.5, -0.7);
\draw[semithick] (2.5, 0).. controls (2.9, 0.3) ..(3.3, 0);
\draw[semithick] (2.5, 0).. controls (2.8, 0.0) ..(3.1, -0.3);
\draw[semithick] (2.5, 0).. controls (2.3, 0.0) ..(2.0, -0.3);
\draw[semithick] (2.5, 0).. controls (2.7, 0.6) ..(3.1, 0.4);
\draw[semithick] (2.5, 0).. controls (2.7, -0.3) ..(2.5, -0.7);
\node at (3, 1.7){$D8$};
\node at (-1.5, 0.8) {$B$};
\node at (2.5, 0.8) {$\overline{B}$};
\node at (0.5, -0.1) {$k$};
\end{tikzpicture}
 \caption{Some strings are connected to D8, some connect vertex and anti-vertex}\label{fig:stringstoD8aninst}
\end{figure}

\item
{\bf All strings are between vertex and anti-vertex}

The last option is -- all strings are long, see Fig.~\ref{fig:stringstoinst}. In this case, the effective electric field is 
\begin{equation}
    E_{eff}=E- NT_{srt}.
\end{equation}

What are the final states in this case? The analogy with the quark-antiquark Schwinger creation
in the confined phase seems to be 
relevant. In that case, we have the string that partially 
lies at the IR wall yielding a similar contribution from the tension. In this case,
the string is broken and we get mesons in the final state.
In the case of the baryon vertex pair creation we expect that the similar breaking of the  $N_c$ strings takes place. Upon the string breaking two options are 
potentially possible. First, the additional vertex-antivertex pair gets created
and we find the pair of glueball-like states involving vertex-antivertex in the final state. These exotic 
glueball-like states
can be in non-singlet state with respect to some global, say, flavor group which 
prevents them from annihilation. Another possible scenario involves the 
reconnection of the strings with the flavor
D8 branes upon the breaking which yields the pair of conventional baryons
which however have the additional contribution to the mass from the $N_c$ 
long strings which undergo the shrinking process later on.

Note that there is an additional statistical suppression factor in this case. Let us fix time 
and consider the $N_c$ strings as the bunch of polymers that starts from the vertex and gets reunioned at the anti-vertex. If the strings are in the antisymmetric state 
then we get the classical problem for the reunion probability for the vicious walkers.
At late times such probability is exponentially suppressed in all dimensions.
In our case, a late time corresponds to the large radius of the circle and we expect 
additional exponential suppression indeed.

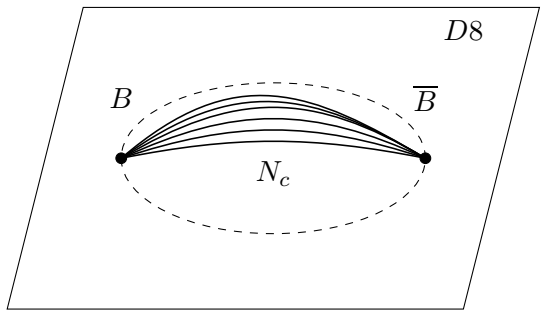
\begin{figure}[h]
\centering
\begin{tikzpicture}
\draw (-3, -2) -- (3, -2) -- (4, 2) -- (-2, 2) -- cycle;
\draw[dashed] (0.5,0) ellipse (2cm and 1cm);
\draw[fill=black] (-1.5, 0) circle (2pt);
\draw[fill=black] (2.5, 0) circle (2pt);
\draw[semithick] (-1.5, 0).. controls (0, 0.5) and (1, 0.5) ..  (2.5, 0);
\draw[semithick] (-1.5, 0).. controls (0, 0.3) and (1, 0.3) ..  (2.5, 0);
\draw[semithick] (-1.5, 0).. controls (0, 0.7) and (1, 0.7) ..  (2.5, 0);
\draw[semithick] (-1.5, 0).. controls (0, 0.9) and (1, 0.9) ..  (2.5, 0);
\draw[semithick] (-1.5, 0).. controls (0, 1.1) and (1, 0.9) ..  (2.5, 0);
\draw[semithick] (-1.5, 0).. controls (0, 1.3) and (1, 0.9) ..  (2.5, 0);
\node at (3, 1.7){$D8$};
\node at (-1.5, 0.8) {$B$};
\node at (2.5, 0.8) {$\overline{B}$};
\node at (0.5, -0.2) {$N_c$};
\end{tikzpicture}
 \caption{All strings are between vertex and anti-vertex}\label{fig:stringstoinst}
\end{figure}

\end{itemize}

Two more remarks are in order. First, if baryon involves the heavy quark the contribution 
of the string connecting the vertex and the corresponding D8 brane has to be taken into account.
This contribution is discussed in Appendix B. Secondly, a more refined analysis \cite{gorsky2012baryon,gorsky2015baryon}
shows that the dyonic instanton solution in 5d gauge theory \cite{lambert1999dyonic} is more relevant for the baryonic state if we take into account the effects of the chiral symmetry breaking on the baryon mass. In the stringy language,
the dyonic instanton solution corresponds to the blow-up of the D4 brane into the D6 brane. In the
field theory language, the additional scalar field is taken into account in the 5d non-Abelian flavor
$SU(2)$ gauge theory, which is briefly mentioned in Appendix A. The corresponding Schwinger process 
for the dyonic instantons is more tricky hence we restrict ourselves to the conventional 5d instantons.

\section{New non-perturbative processes in electric field}
\subsection{Spontaneous decay of a neutron in electric field}

\begin{figure}[b]
\centering
\begin{minipage}{4cm}
\begin{tikzpicture} [> = stealth]
\draw[thick] (0, -2) -- (0, -1);
\draw[thick] (0, 1) -- (0, 2);
\draw[thick, ->- = 0.5] (0, -1) arc (-20:20:2.95);
\draw[thick, ->- = 0.5] (0, 1) arc (90:270:1);
\node at (0.2, -1.5){$n$};
\node at (0.2, 1.5){$n$};
\node at (0.4, 0.2) {$p$};
\node at (-1.3, 0.2) {$\pi^-$};
\node at (-1.5, 2) {$(a)$};
\end{tikzpicture}	
\end{minipage}
\hspace{1.0cm}
\begin{minipage}{4cm}
\begin{tikzpicture} [> = latex]
\draw[thick] (0, -2) -- (0, -1);
\draw[thick] (0, 1) -- (0, 2); 
\draw[thick, ->- =0.5] (0, -1) arc (-60:60:1.16);
\draw[thick, ->- = 0.5] (0, 1) arc (100:260:1.02);
\node at (0.2, -1.5){$n$};
\node at (0.2, 1.5){$n$};
\node at (0.8, 0.2) {$p$};
\node at (-1.1, 0.2) {$\rho^-$};
\node at (-1.5, 2) {$(b)$};
\end{tikzpicture}
\end{minipage} \caption{Bounce for the neutron decay: (a) via $\pi^-$ channel; (b) via $\rho^-$ channel}\label{pic1}
\end{figure}
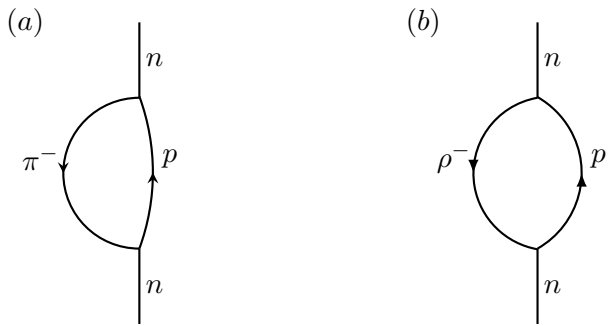

In this Section, we shall consider the new non-perturbative processes involving
the composite worldline instantons.
Consider first the new neutron decay channel 
in the external electric field. 
Recall that the standard process of neutron  $\beta$-decay proceeds 
perturbatively via the weak interaction and virtual W-boson. We suggest
the non-perturbative process which gets started from the composite 
worldline instanton, see Fig.~\ref{pic1} and weak interaction enters the game at the 
second stage. In the first stage the neutron decays into a proton 
and charged meson, $\pi^{-}$-meson or $\rho^{-}$ meson,
while in the second stage, the charged meson weakly decays into the corresponding final state.

This new decay channel is somewhat similar to the 
non-perturbative decay of electrically charged particles in the magnetic
field and magnetic monopole in the electric field 
discussed in \cite{gorsky2002schwinger,monin2005monopole}.
The mechanism is similar to the Schwinger-like pair production but instead of the 
symmetric pair in the final state we have asymmetric final states. Say, for the
decay of the electrically charged particle in the magnetic field, we get the
monopole and dyon in the final state \cite{gorsky2002schwinger}. The corresponding
bounce solution is an asymmetric composite curve contrary to the circular worldline instanton
for the symmetric final state.
Note that in $(1+1)$ the  Schwinger process with the particle in its initial state is similar to 
the induced
false vacuum decay first considered in \cite{affleck1979induced}. Another example 
of the similar process in $(1+1)$ is the non-perturbative decay of the bound state in the 
Thirring model in \cite{gorsky2006particle}. The preexponential factor for the
monopole decay has been evaluated in \cite{monin2007nonperturbative} and the 
account of the temperature was taken in \cite{monin2008semiclassical}.

Technically we are looking for the non-perturbative correction  to the neutron propagator which
is responsible for the decay process. The relevant correction to the propagator of the neutron $G_N$ 
from $(0,0,0,0)$ to $(0,0,0,T)$,  where T is Euclidean time, reads  as
follows 
\begin{equation}
\label{correction}
   \delta G_N(0,T)= g_{\pi NN}^2\int G_N(z,0)G_N(T,w) Tr[G_P(w,z,E)G_{\pi}(w,z,E)]dz dw
\end{equation}
where $G_P(z,w,E)$ is propagator of proton in the external electric field while 
$G_{\pi}(z,w,E)$ is propagator of $\pi^{-}$ in the external electric field. The similar
expression takes place if we consider $\rho^{-}$-meson trajectory.
The explicit expressions for the propagators in a constant homogeneous electric
field are known. We consider a neutron at rest and 
take use of the $np\pi^{-}$ vertex  or $np\rho^{-}$ vertex \cite{machleidt1989meson} 
\begin{equation}
    L_{\pi NN}=\bar{N}[F_{\pi}^{-1}g_A \vec{t}(\vec{\sigma}\vec{\nabla},)\vec{\pi}]N
\end{equation}
\begin{equation}
    L_{\rho N N}=\bar{N}[F_{\pi}^{-1}g_V \vec{t}\gamma_{\mu}\vec{\rho_{\mu}}]N,
\end{equation}
where axial coupling $g_A$ and vector constant $g_V$ are experimentally known.

In the approximation when the Larmour radius of the meson trajectory in the
$(x_3,x_0)$ plane is large the saddle-point approximation in (\ref{correction}) is available 
\cite{gorsky2002schwinger,monin2005monopole,monin2007nonperturbative} and the emerging picture is quite transparent.
The two charged particles propagate
in the Euclidean time along circle segments trajectories in the external electric field which
in the Euclidean time plays the role of magnetic field.  Therefore the worldline instanton 
involves two intersecting segments of different radii defined by the masses of proton
and meson. The equilibrium condition at the point of intersection is found by the
total zero force condition defined by the particle masses.

The action evaluated at the saddle point  world-line instanton solution reads as
\begin{equation}
    S_{inst}=m_{\pi}L_{\pi} + M_p L_p - eE(Area) - M_n H,
\end{equation}
where $L_{\pi}, L_{p}$ are the lengths of the corresponding segments of 
proton and meson trajectories and $H$ is the distance between the junction points in  Euclidean time $t_E$.
Since the interaction vertex involves the momentum the action 
has to be multiplied by the corresponding factor derived from 
the equilibrium condition at the vertex. It contributes to the preexponential factor.
In the leading approximation taking into account that $m_{\pi}<< M_p$ the action 
can be approximated by 
\begin{equation}
    S_{inst}^0\propto \frac{\pi m_{\pi}^2}{2eE}
\end{equation}
The correction to the neutron propagator can be considered as the  non-perturbative  imaginary part of the neutron mass
\begin{equation}
    Im\delta M_{n}\propto e^{-S_{inst}}.
\end{equation}

More accurate saddle-point action for  the composite 
worldline instanton  when the
$\rho$-meson propagates  in the composite worldline instanton
can be derived similar to \cite{monin2005monopole} and reads as
\begin{equation}
    S_{inst}= \frac{m_{\rho}^2}{eE} \arccos{\frac{M_n^2 +m_{\rho}^2 -M_p^2}{2m_{\rho}M_n}} + 
    \frac{M_{p}^2}{eE}\arccos{\frac{M_n^2 -m_{\rho}^2 + M_p^2}{2M_pM_n}}-
    \frac{m_{\rho}M_n}{eE}\sqrt{ 1 - \left(\frac{M_n^2 +m_{\rho}^2 -M_p^2}{2m_{\rho}M_n}\right)^2}
\end{equation}
For the $\rho$-meson decay channel the instanton configuration is closer
to the circle.
Certainly, this expression is subject to the corrections via meson exchanges in the loop
which generate the interaction between the neutron and meson in the loop. However 
to get the controllable approximation we assume that such interactions
inside the circle are suppressed exponentially due to the meson masses.

It is instructive to compare this non-perturbative process with another non-perturbative
process - scattering of monopole pair into the dyon-antidyon pair \cite{atiyah1985low}. The process can be interpreted
in two ways. First, it can be treated as the exchange by the massive charged particle
which yields the factor $\exp(-MR)$ where R is some characteristic scale to be determined. The more 
rigorous description of low-energy scattering involves the geodesic motion on the 
moduli space of a 2-monopole solution. The non-perturbative Atyah-Hitchin metric 
for two monopole  moduli space has been found in \cite{atiyah1985low}
and in some limit it gets reduced to Taub-NUT geometry \cite{manton1985monopole} which 
allows very transparent semiclassical interpretation. There is the particular geodesic in the
Atyah-Hitchin metric when the motion along the internal $S^1$ corresponding to electric charge 
gets induced. It is this geodesic which corresponds to the monopoles into dyons process. 

Our bounce picture corresponds to the first description. 
The meson propagator yields the $\exp(-MR)$ factor
while the scale $R$ is fixed by the saddle point bounce configuration.
However, we expect that the second approach based on the geodesics
on the moduli space of the Skyrmion works as well. To this aim,
one has to get the modification of the metrics due to the meson mass
and the electric field. We plan to discuss this approach elsewhere.

\subsection{Production of $p\bar{n}\pi^{-}$  in the electric field}

\begin{figure}
    \centering
    \begin{tikzpicture}[>=stealth]
    \draw[thick] (0, - 0.8) -- (0, 0.8);
    \draw[thick, ->- = 0.5] (0, 0.8) arc (110:-110:0.85);
    \draw[thick, ->- = 0.5] (0, -0.8) arc (335:25:1.90);
    \node at (-0.3, 0.1){$\overline{n}$};
    \node at (-3.3, 0) {$p$};
    \node at (1.5, 0) {$\pi^-$};
    \end{tikzpicture}
    \caption{The Euclidean trajectory for $p\bar{n}\pi^{-}$ creation}
    \label{fig:pnpi_creation}
\end{figure}
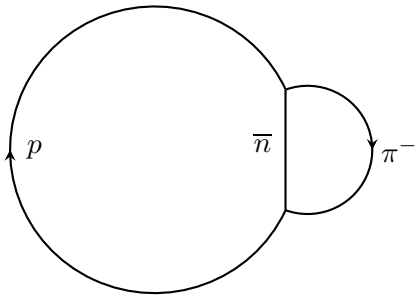
There is another interesting non-perturbative process in an electric field without particles
in the initial state. Consider the following composite worldline instanton:
the proton-antiproton trajectory gets started at some initial point, there is
the junction at which proton trajectory gets glued with the neutron and $\pi^{+}$
trajectories, see Fig~\ref{fig:pnpi_creation}. The equilibrium condition is satisfied at a junction point. 
The neutron is neutral hence its trajectory is the straight line. After the 
second junction point the proton and antiproton trajectories get connected. 
The action at this composite worldline instanton is evaluated as before in 
the weak external field approximation. 
In the case of $\pi^{-}$ involved into the composite instanton the deformed segment 
of the  circle 
is small hence the probability in the leading approximation has the form
\begin{equation}
    \omega \propto \exp(-\frac{\pi M_p^2}{eE}).
\end{equation}
If the $\rho^{-}$ is involved in the composite instanton the trajectory 
is modified considerably and the probability reads as
\begin{multline}
    \omega \propto \exp\left(-\frac{\pi (M_p^2 + m_\rho^2)}{e E} - \frac{M_p M_n}{eE} \sqrt{1 - \left( \frac{M_p^2 - m_\rho^2 + M_n^2}{2 M_p M_n}\right)^2} + \right. \\+ \left. \frac{M_p^2}{e E} \arccos \frac{M_p^2 -m_\rho^2 + M_n^2}{2 M_p M_n} +\frac{m_\rho^2}{eE} \arccos \frac{m_\rho^2 - M_p^2 + M_n^2}{2 m_\rho M_n} \right).
\end{multline}

At $t_E=0$ the solution gets rotated from the Euclidean to Minkowski space-time.
The final state involves the antiproton, neutron, and meson with the positive charge.
The analogous process with the same probability when the junctions take place at the antiproton trajectory yields
in the final state the proton, antineutron, and meson with negative charge.
The evolution in the Minkowski state involves the motion of two charged
particles in opposite directions in the electric field and a neutron at rest. The entanglement of three
final states deserves special study.

Note that the bounce solution is rotationally asymmetric and we can for instance 
rotate it at $90^{o}$. In this case, we have two options to make the analytic continuation
to the Minkowski space. In the first case, we get the subleading correction to the
proton-antiproton pair production while in the second case, we get the subleading 
correction to the meson pair production. All amplitudes related by the 
rotation and different analytic continuations are expected to be related
by the unitarity similarly to the discussion in \cite{gorsky2007spontaneous}.

\subsection{Does proton decay in an electric field?}

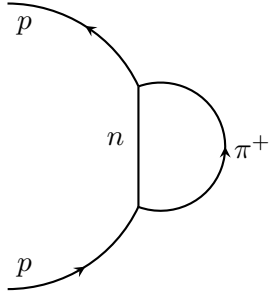
\begin{figure}
\centering
    \begin{tikzpicture}[>=stealth]
    \draw[thick] (0, - 0.8) -- (0, 0.8);
    \draw[thick, ->- = 0.5] (0, -0.8) arc (-110:110:0.85);
    \draw[thick, ->- = 0.5] (-1.72, -1.89) arc (-90:-25:1.90);
    \draw[thick, ->- = 0.5] (0, 0.8) arc (25:90:1.90);
    \node at (-0.3, 0.1){$n$};
    \node at (-1.5, 1.6) {$p$};
    \node at (-1.5, -1.6) {$p$};
    \node at (1.5, 0) {$\pi^+$};
    \end{tikzpicture}    \caption{Conjectual Euclidean trajectory yielding the proton decay in electric field}
    \label{fig:pdecay}
\end{figure}

One could question if there is proton decay into a neutron and $\pi^{+}$-meson in external electric field.
Formally the question concerns the 
existence of the non-perturbative imaginary part in the proton propagator in the external 
electric field at the second order in the couplings $g_V$ or $g_A$.
Hence we have to evaluate the following contribution to the proton 
propagator in the electric field
\begin{equation}
\label{proton}
      \delta G_p(0,T,E)= g_{\pi NN}^2\int G_p(z,0,E)G_p(T,w,E) Tr[G_n(w,z)G_{\pi}(w,z,E)]dz dw
\end{equation}
The situation is a bit different in comparison with the neutron decay 
in the electric field since the particle in the initial state is charged, see Fig.~\ref{fig:pdecay}.
The potential tunneling process involves three time intervals $Minkowski \rightarrow Euclid \rightarrow Minkowski$
hence we have to perform the analytic continuation twice. For the neutron's initial state the
first analytic continuation from Minkowski to Euclidean space is simple however for proton it is not the case
due to its accelerating motion in an electric field. 

If we consider the proton at rest in the accelerating
frame the effective Rindler metric gets induced. Hence upon the Wick rotation,
we should look for the composite worldline instanton in the Euclidean Rindler space in the 
external magnetic field which follows from the electric field in the Minkowski space. There are some subtle points
in gluing the neutron and $\pi^{+}$-meson trajectories in the Euclidean Rindler space because of its
wedge structure. On the other hand, the formal evaluation of the imaginary 
part in (\ref{proton}) in the saddle point approximation also is questionable since there are some 
additional singularities and a selection of the proper contour requires additional analysis. We postpone the question concerning this process for the separate study.

\section{Conclusion}

In this study, we have discussed the  Schwinger production of baryons in the
constant electric field in the worldline instanton approach. A baryon in the confinement phase is considered 
as instanton in the flavor group in the holographic framework or as Skyrmion in the ChPT.
The circular circle saddle point configuration for the baryon trajectory has been 
found.
At the weak external field in ChPT, the radius of the circle is large hence we can neglect the 
corrections due to the meson exchanges and the main contribution comes from the classical trajectory
of the Skyrmion in the Euclidean space-time. In the holographic QCD the production of the 
pair of baryon vertexes has been considered and we have mentioned some subtle points 
due to the strings attached to the baryonic vertex.
We have suggested the new rare exponentially suppressed processes 
in the external electric field and the corresponding probability rates have been evaluated in the 
leading approximation. Presumably, such processes  like decay of a neutron can be of 
some importance in the early Universe.

There are several issues for further study.
It would be interesting to evaluate the effects of the meson exchanges to the probability rate and the effects 
of the multiple S-Skyrmions.
It is also natural to question if the pair of 
pancake baryons for $N_f=1$ can be created in the external electric field. In this case, we
have a torus-like bounce with the meson currents at the boundary. Somewhat similar 
configurations are expected if we consider the dyonic instanton realization of
baryons in holographic QCD. The analysis of this study certainly can 
be generalized for a case of inhomogeneous electric field.

The new  channels like  neutron decay are exponentially suppressed and therefore beyond
the experimental availability.
We could
question if some external factors could enhance the probability rates. 
Usually, the induced Schwinger processes are considered at finite temperature \cite{gould2017thermal}
or as photon stimulated process \cite{dunne2009catalysis,monin2010semiclassical,monin2010photon}. The key question concerns the mechanism of transition 
of the energy of the external field to the collective degrees of freedom describing 
the bounce solution. Since the main contribution to the rate probability comes
from the circular trajectory of the baryonic vertex the main effect of the
temperature comes from the deformation of the vertex euclidean trajectory.
The vertex trajectory gets modified strongly when the diameter of the circle
coincides with the inverse temperature $2R_{b}=\frac{2m_p}{eE}=T^{-1}$. 
At larger temperatures, the bounce configuration gets modified 
and acquires the fish-like form.

The photon stimulated Schwinger proton production can be evaluated similarly to \cite{dunne2009catalysis,monin2010semiclassical,monin2010photon} when 
the bounce configuration is modified as well. If we would consider the photon stimulated processes 
the new difficulty influences the analysis. The external photon now can interact with
the anomalous term in the electric current hence the saddle point configuration 
should be reconsidered.

Another interesting question concerns the entanglement of the particles in the final state.
This question has been discussed first in \cite{jensen2013holographic,sonner2013holographic} and the detailed analysis in the confinement phase has been done in \cite{grieninger2023entanglement,grieninger2023entanglement2}. It would be interesting 
to consider these issues for all new processes suggested in this study.

We are grateful to S. Kulagin for the useful discussion and to A. Vainshtein and D. Kharzeev  for the
comments on the manuscript.
A.G. thanks Nordita and IHES where the parts of this work have been done for the 
hospitality and support.

\section{Appendix A. Dyonic instanton}
For completeness let us recall the dyonic instanton solution 
providing a more  general holographic representation of baryon 
\cite{gorsky2012baryon,gorsky2015baryon}. It takes into account 
the pseudoscalar field in the bifundamental representation of left and right
flavor groups. The chiral condensate contributes considerably in this
representation to the baryon mass \cite{gorsky2015baryon}. In the stringy terms the wrapped D4 brane gets expanded into the D6 brane.

The dyonic instanton in  the $SU(2)$ YM theory in 5d with a scalar field in adjoint representation
has been found in \cite{lambert1999dyonic}. 
For simplicity, we assume that the scalar field has no potential and its vev is arbitrary.
The model action in 
$$
S = \int d^5 x \Tr \left(\dfrac{1}{2}F_{\mu\nu}^2 + \left(D \phi \right)^2 \right)
$$
We use covariant derivative $D_\mu \phi = \partial_\mu \phi - i g [A_\mu ,\phi]$ and field strength definition
$$
F_{\mu \nu} = \partial _\mu A_\nu - \partial_\nu A_\mu - i g[A_\mu, A_\nu].
$$
We consider an instanton solution, extended in some spatial direction. We choose the direction $x_5$ and separate $1,\dots ,4$ field component and $5$ component:
$$
S = \int d^5 x \Tr \left( \frac{1}{2}F_{\mu \nu} ^2 + E_\mu^2 + (D_\mu \phi)^2 + (D_5 \phi)^2\right)
$$
We define electric field as $E_\mu = F_{\mu 5}$, $\mu = 1, \dots , 4$

We proceed in a way similar to dyonic instanton in real space, extracting the full squares in action
$$
S = \int d^5 x \Tr \left( \dfrac{1}{4} (F_{\mu \nu} - * F_{\mu \nu})^2 + (D_\mu \phi - E_\mu)^2 + (D_5 \varphi)^2 + \dfrac{1}{2} F_{\mu \nu} * F_{\mu \nu} + 2E_\mu D_\mu \phi \right)
$$
The last two terms are full derivatives and their contribution to energy depends only on boundary conditions. $F*F$ integral is proportional to instanton charge and $E_\mu  D_\mu \phi$ is proportional to electric charge.
Therefore to minimize energy per unit length in $x^5$ direction first two brackets should be equal to zero:
$$
F_{\mu \nu} = * F_{\mu \nu}, \quad E_\mu = D_\mu \phi.
$$

The first equation is solved by  instanton field, which in singular gauge is
$$
A_\mu = \frac{2}{g} \frac{\rho^2}{r^2(\rho^2 +r^2)} \overline{\eta}^a_{\mu \nu} x_\nu \frac{\sigma^a}{2}.
$$

Then we can find field $\phi$ configuration by solving the equation
$$
D^2 \phi = 0.
$$
The answer for zero mode is known
$$
\phi = v \frac{r^2}{r^2 + \rho^2} \frac{\sigma^3}{2}.
$$
We assume that scalar vev is $v \sigma^3/2$.
The corresponding electric field is obtained by direct differentiation. For example
$$
E_1 = \left(
\begin{array}{cc}
 -\frac{\rho ^2 v x_1}{\left(\rho ^2+x_1^2+x_2^2+x_3^2+x_4^2\right){}^2} & \frac{\rho ^2 v \left(x_3-i x_4\right)}{\left(\rho ^2+x_1^2+x_2^2+x_3^2+x_4^2\right){}^2} \\
 \frac{\rho ^2 v \left(x_3+i x_4\right)}{\left(\rho ^2+x_1^2+x_2^2+x_3^2+x_4^2\right){}^2} & \frac{\rho ^2 v x_1}{\left(\rho ^2+x_1^2+x_2^2+x_3^2+x_4^2\right){}^2} \\
\end{array}
\right).
$$
This electric field corresponds to potential $A_5 = \phi$, therefore $D_5 \varphi = 0$ (fields do not depend on $x_5$).

We can define $U(1)$ electric field as $\mathcal{E}_\mu = \dfrac{1}{v}\Tr (E_\mu \phi)$, thus electric charge
$$
q = \int dS^\mu \mathcal{E}_\mu = 2 \pi^2 \rho^2 v. 
$$

Now we consider the energy-momentum tensor for this field configuration.
For gauge field
$$
T^{\mu \nu} = \Tr \left( F^{\mu \alpha} F^{\nu \alpha} - \frac{1}{4}\delta^{\mu \nu} F_{\alpha \beta}^2\right) .
$$
For scalar field 
$$
T^{\mu \nu} = \Tr \left( 2 D_\mu \phi D_\nu \phi - \delta^{\mu \nu} (D_\alpha \phi)^2\right).
$$
We are interested in $\mu 5$ components (energy flow in the direction of the instanton). The scalar field contribution is zero, $D_5 \phi = 0$.
For the gauge field we have 
$$
T^{\mu 5} = \Tr \left( F^{\mu \alpha} F^{5 \alpha} \right) = \Tr (F_{\mu \alpha} E_\alpha).
$$ 
By direct calculation we have
\begin{align*}
T^{15}& = \frac{12 v \rho^4 x_2 }{(\rho^2 + r^2)^4}  ,\quad T^{25} = - \frac{12 v \rho^4 x_1 }{(\rho^2 + r^2)^4}\\
T^{35}& =-  \frac{12 v \rho^4 x_4 }{(\rho^2 + r^2)^4}  ,\quad T^{45} = \frac{12 v \rho^4 x_3 }{(\rho^2 + r^2)^4}.
\end{align*}
Now we can calculate the rotation generator for our solution
$$
L_{\mu \nu} = \int d^4 x (x_\mu T_{5 \nu} - x_\nu T_{5 \mu})
$$

$$
L_{12} = - L_{21} = \int d^4 x \frac{(-x_1^2 - x_2^2) 12 v \rho^4}{(\rho^2 + r^2)} = - 2\pi^2 \rho^2 v.
$$
Similarly 
$$
L_{34} =- L_{43}= 2 \pi^2 \rho^2 v.
$$
Other components are equal to zero. Thus we find that the solution is stabilized by its charge or angular momentum. 
Angular momentum is proportional to the electric charge, therefore this is the same mechanism.

\section{Appendix B. Baryon with a heavy quark}

In this Appendix, we will consider the Schwinger effect for baryon involving a heavy quark. In the WSS model, the quark mass can be described as the separation between the flavor D8 branes. One of the strings connects the instanton with the separated D8 brane located at large radial coordinate $u = u_f$. We should add to the action contribution from NG action from the string ending at that brane. This action is proportional to the minimal surface area connecting the instanton trajectory and the D8 brane. 

The full action for the Schwinger process is
\begin{equation}
    S = 2\pi M r_0 - q E r_0^2 + S_{NG},
\end{equation}
where $S_{NG}$ is the contribution from the string, attached to the separated brane. The string action is proportional to the worldsheet area, where the worldsheet is embedded in the WSS metric
\begin{equation}
    ds^2 = \left(\frac{u}{R}\right)^{3/2} (\eta_{\mu \nu }dx^\mu dx^\nu + f(u) d\tau^2) + \left(\frac{R}{u} \right)^{3/2}
\left(\frac{du^2}{f(u)} + u^2 d\Omega_4^2 \right), \quad f(u) = 1 - \frac{u_k^3}{u^3}.
\end{equation}

The instantonic circle sits at the top of the "cigar" where $u = u_k$.
We assume that the surface is rotationally symmetric and therefore can be parametrized by a single function $r = r(u)$ that describes the surface radius at coordinate $u$.
The induced metric  on the surface is
\begin{equation}
ds^2 =\left( \frac{u}{R} \right)^{3/2} (r^{\prime 2}du^2 + r^2 d\varphi^2 ) + \frac{R^{3/2}du^2}{u^{3/2} f(u)},\end{equation}
where $\varphi$ is an angle coordinate.
Therefore the Nambu-Goto action is
\begin{equation}
S = T_{s} \int d^2 x \sqrt{g} = T_S \int d\varphi du\, r \left(\frac{u}{R} \right)^{3/2} \sqrt{r^{\prime 2} + \frac{R^3}{u^3 - u_k^3}},
\end{equation}
and the corresponding equation of motion reads as
\begin{equation}
 \frac{r r'' }{1 + r^{'2}(u^3 -u _k^3)/R^3} + \frac{3  r r'}{2u } + \left[\frac{3 r r' u^2 /(2R^3)}{1 + r^{\prime 2} (u^3 - u_k^3)/R^3} -  1 \right] \frac{R^3}{u^3 - u_k^3} = 0.
\end{equation}

At $u = u_k$ the surface is attached to the instanton, so the boundary condition is $r(u_k) = r_0$, and the other condition is fixed by the regularity of the solution at $u = u_k$:
\begin{equation}
r(u_k) = r_0 , \quad r'(u_k) = \frac{2R^2}{3 r_0 u_k}.
\end{equation}
The solution should end at value $u = u_q \gg u_k$ proportional to the heavy quark mass.
In the limit of large $u_q \gg u_k$, we can use large $u$ asymptotic 
\begin{equation}
r(u) = r_\infty - \frac{c}{u} + O\left( \frac{1}{u^2}\right), \quad c = \frac{R^3}{r_{\infty}}, \quad u \to \infty.
\end{equation}

Corresponding expansion for action is 
\begin{equation}
S_{NG} \approx 2\pi T_s \int_{u_k}^{u_q} du\left(r_\infty - \frac{R^2}{2 r_\infty u} \right) \approx 2 \pi T_s \left(r_{\infty} (u_k - u_s) - \frac{R^3 }{2 r_\infty } \ln \left(\frac{\alpha u_q}{u_k} \right)\right).
\end{equation}
where $\alpha$ is some dimensionless number depending on $r_0$. The value of $r_\infty$ can be found only numerically, the calculations show that $r_\infty - r_0 \sim 1/r_0$, so in the limit of large $r_0$ (and weak field) we can set $r_\infty \approx r_0$, and our estimation for action is
\begin{equation}
S_{NG} = 2 \pi T_s (u_q - u_k) r_0 - \frac{2 \pi T_s R^3 }{2 r_0} \ln \left(\frac{\alpha u_q}{u_k} \right).
\end{equation}
The full action is
\begin{equation}
S = 2 \pi M_1 r_0 - \pi q E r_0^2 - \frac{A}{r_0}, \quad A =  \pi T_s R^3 \ln \left(\frac{\alpha u_q}{u_k} \right).
\end{equation} 
Here $M_1 = M +T_s (u_q - u_k) $ is baryon mass with heavy quark contribution.
We have to minimize this action with respect to $r_0$.  In the weak field limit $qE \ll M_1^2$ the circle radius $r_0$ is large and the last term is a small correction. Therefore in the leading approximation the radius value is the same as in the case without the NG action but with corrected baryon mass, $r_0 = M_1/qE$.
\begin{equation}S = \frac{\pi M_1^2 }{q E} - A\frac{q E}{M_1}.
\end{equation}
Therefore in a weak field limit the heavy quark yields an additional contribution to the baryon mass and a term in action, linear in $q E$.
The Schwinger process probability is  
\begin{equation}
w \sim e^{-S}  = \exp \left(- \frac{\pi M_1^2 }{q E} + A \frac{q E}{M_1} \right),
\end{equation}
We can express the parameters of the WSS model \cite{sakai2005low} as
\begin{equation}
T_s = \frac{1}{2\pi l_s^2}, \quad R^3 = \pi g_s N_c l_s^3, \quad g_s = \frac{g_{YM}^2}{2\pi M_{KK} l_s}
\end{equation}
where $l_s$ is string length, $g_s$ is string coupling constant,  $N_c$ is number of colors, $M_{kk}$ is Kaluza-Klein mass and $g_{YM}$ is 4d Yand-Mills coupling constant. Therefore final expression for the constant $A$ is
\begin{equation}
    A = \frac{g_{YM}^2 N_c}{4 \pi M_{kk}} \ln \left(\frac{\alpha u_q}{u_k} \right).
\end{equation}

\bibliography{references.bib}

\end{document}